\documentclass[multphys,vecphys]{svmult}

\usepackage{makeidx}         
\usepackage{graphicx}        
\usepackage{multicol}        
\usepackage[bottom]{footmisc}

\makeindex             

\newcommand{\hi}{\mbox{H{\sc i}}}
\newcommand{\hii}{\mbox{H{\textsc{ii}}}}
\newcommand{\ha}{H$\alpha$}
\newcommand{\kms}{km s$^{-1}$}

\begin{document}

\title*{Harmonic analysis of the \ha\ velocity field of NGC 4254}
\author{Laurent Chemin\inst{1,2}, Olivier Hernandez\inst{1},
Chantal Balkowski\inst{2},  Claude Carignan\inst{1} \& Philippe Amram\inst{3}}
 \authorrunning{Chemin et al.}  
 
\institute{Universit\'e de Montr\'eal, D\'ept. de Physique, 
C.P. 6128 succ. centre-ville, Montr\'eal (QC), CANADA H3C 3J7
\texttt{chemin@astro.umontreal.ca}
\and Observatoire de 
Paris, section Meudon, GEPI, CNRS-UMR 8111 \& Universit\'e Paris 7, 5 Pl. Janssen, 92195, Meudon, France \and 
Observatoire Astronomique Marseille-Provence, 2 Pl. Le Verrier, 13248, Marseille Cedex 4, France}
%
%
\maketitle

The ionized gas kinematics of the  Virgo Cluster galaxy NGC 4254 (Messier 99) is analyzed by 
an harmonic decomposition of the velocity field into Fourier coefficients. 
The  aims of this study are to measure the kinematical asymmetries of Virgo cluster galaxies 
 and to connect them to the environment.  The analysis reveals significant  $m=1,2,4$ terms which 
 origins are discussed. 

\section{Introduction}
Galaxies in clusters are sensitive to environmental effects like the cluster tidal field, 
gravitational encounters with other galaxies, galaxy mergers,  ram pressure stripping and  
accretion of gas (see e.g. Moore et al. 1998, Vollmer et al. 2001). 
Such external events dramatically affect their structure,  
triggering internal  perturbations  like bars or oval distorsions (e.g. Bournaud \& Combes 2002), spirals,  warps 
(Huang \& Carlberg 1997)  or lopsidedness (Bournaud et al. 2005). Their kinematics
is also disturbed, as revealed by long-slit spectroscopy 1-D rotation curves (Rubin et al. 1999).
 
\begin{figure}[!t]
\centering
\includegraphics[height=3cm]{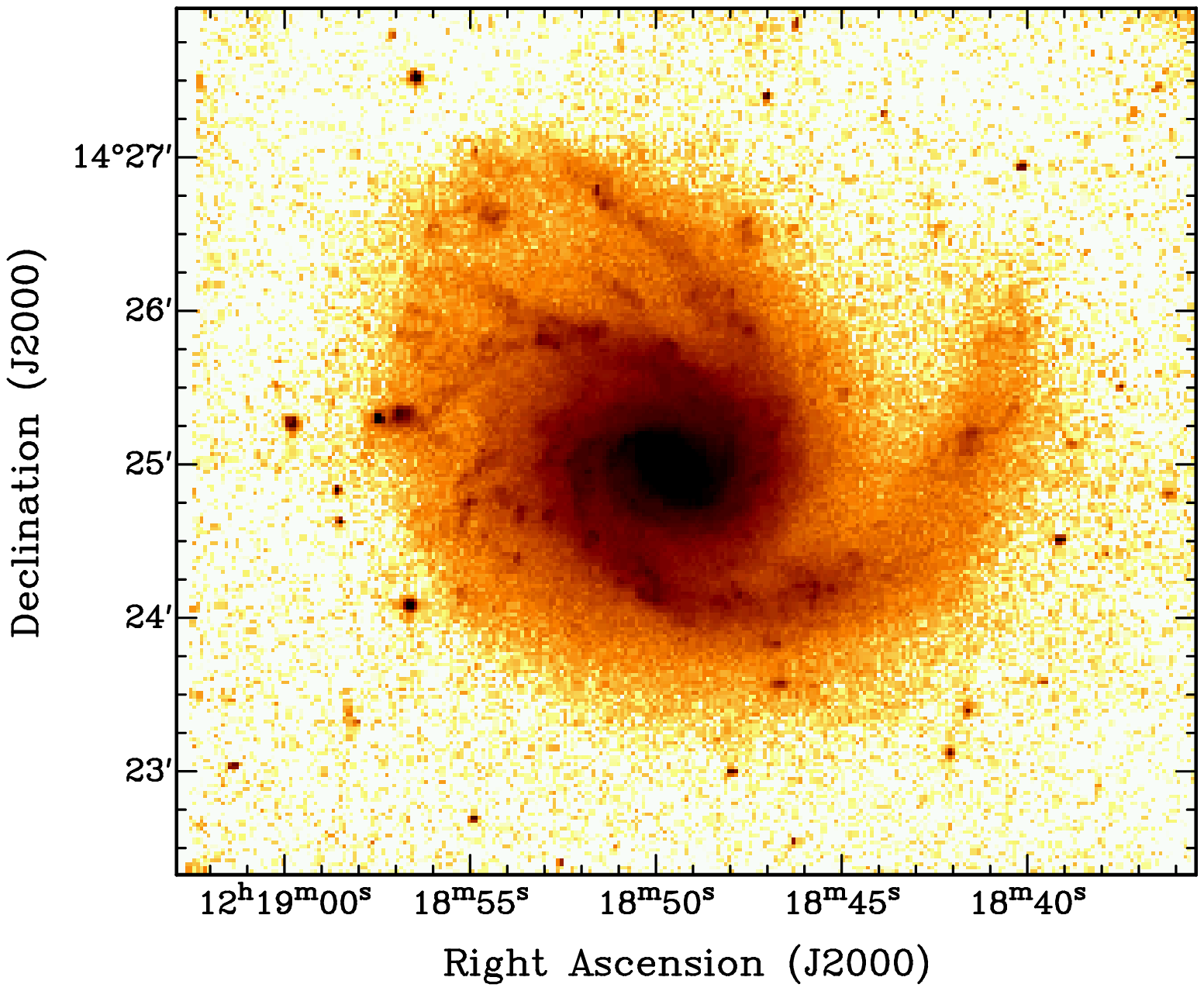}\hspace*{-0.5cm}\includegraphics[height=3cm]{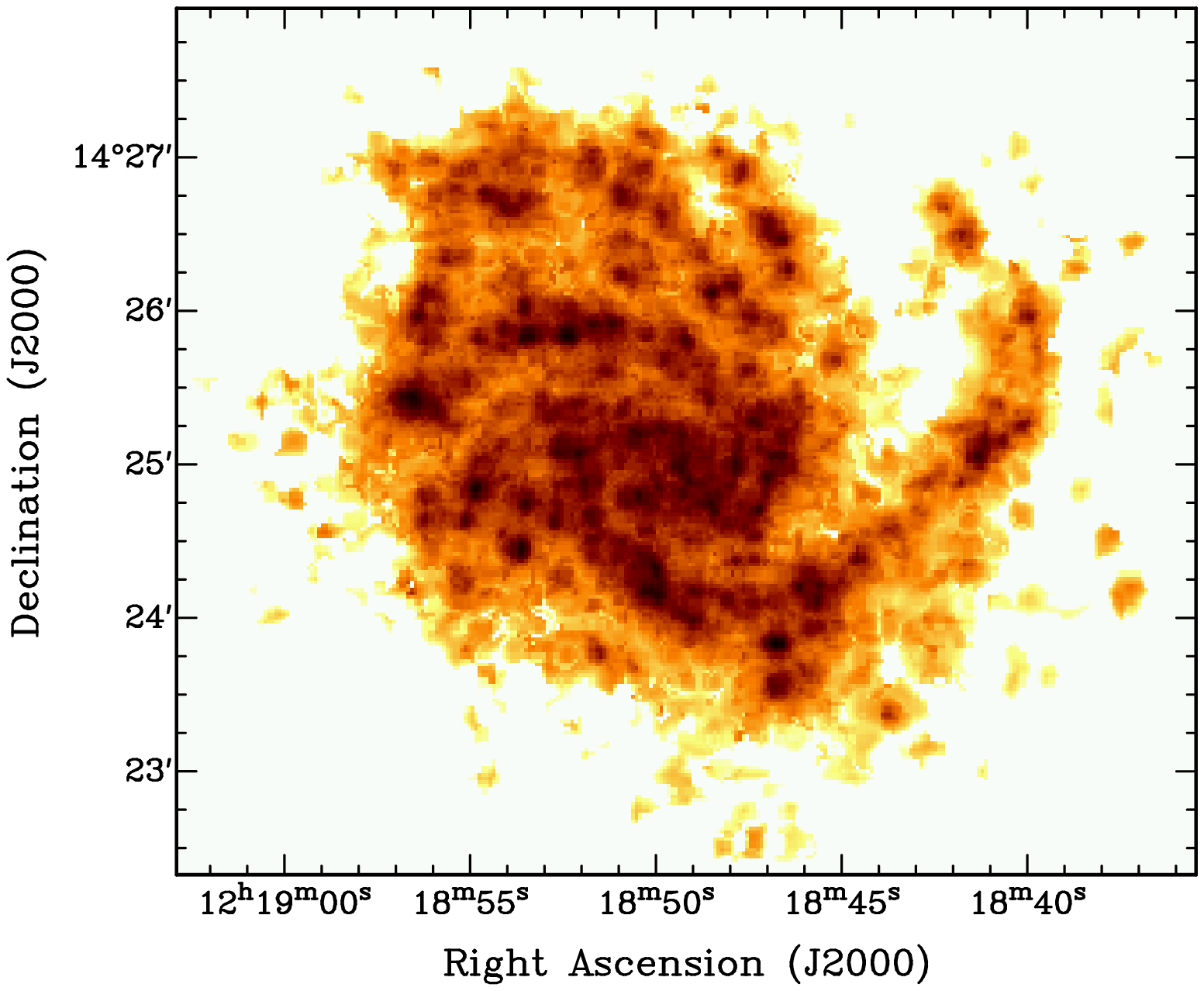}\hspace*{-0.5cm}\includegraphics[height=3cm]{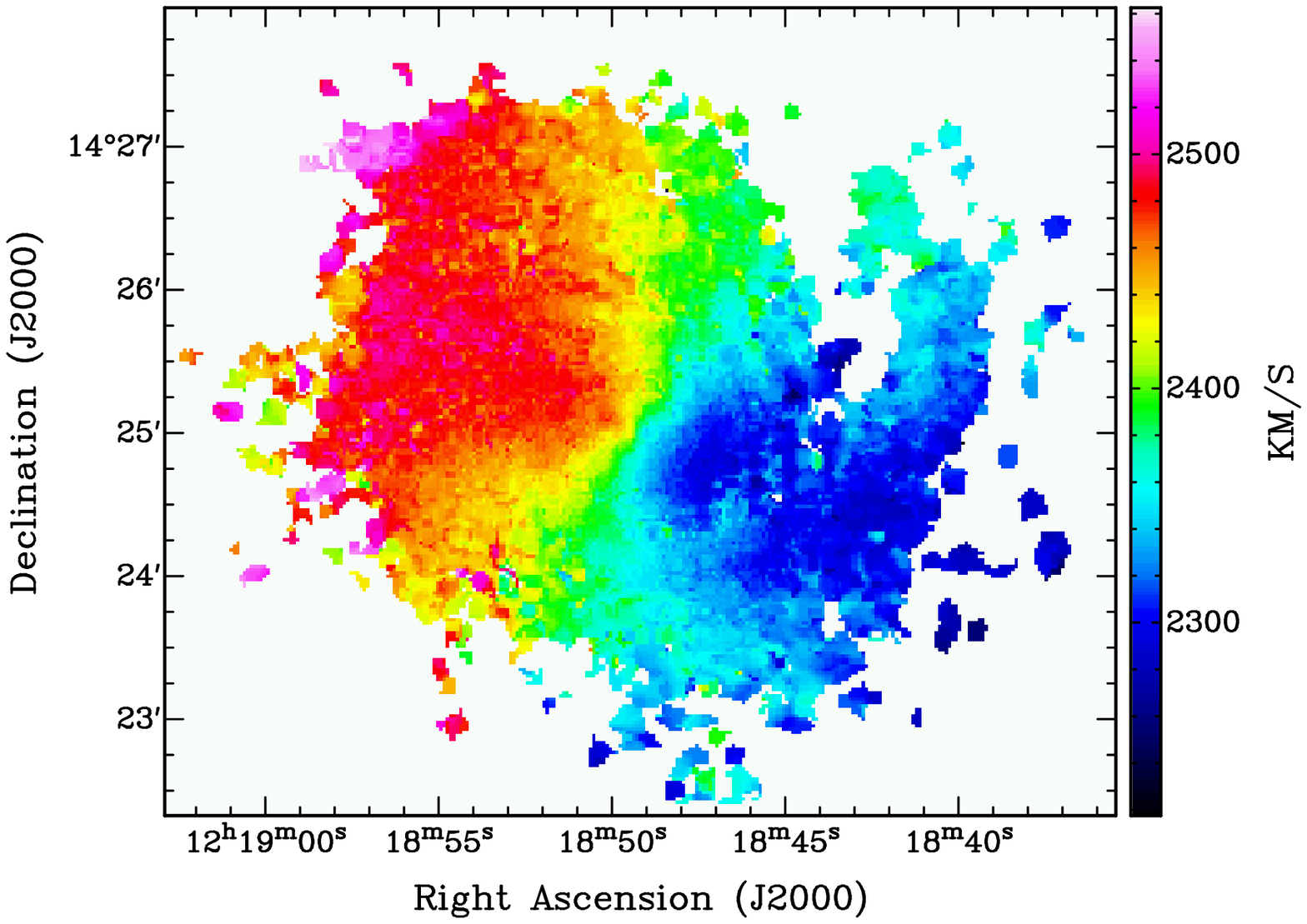}
\caption{ $H-$band image, \ha\ emission and velocity maps of NGC 4254. The NIR image is from Gavazzi et al. (2003).}
\label{fig1}       
\end{figure}

 High-resolution \ha\ velocity fields were obtained for 30 Virgo cluster galaxies (Chemin et al. 2005)
in order to study the  degree of perturbation of their 2-D kinematics and the influence of the 
environment on the kinematics. The harmonic analysis is a powerful tool to detect kinematical 
anomalies, as already shown on \hi\ velocity fields (Schoenmakers, Franx \& de Zeeuw 1997). 
This technique is applied to the \ha\ velocity field of NGC 4254 (Figure~\ref{fig1}). 

\section{The Virgo Atlas : observations}
\label{sec:1}

The data acquisition and reduction are   described in details in Chemin et al. (2005). Data cubes of 
30 galaxies were obtained in the \ha\ emission-line at  the Observatoire du mont M\'egantic (Canada), 
the   Observatoire de Haute-Provence, the ESO 3.6-m telescope and the  Canada-France-Hawaii-Telescope
 between 2000 and 2005. The instrument is composed of a focal reducer and a Fabry-Perot interferometer coupled with a photon-counting system.
  The spectral sampling of the observations varies between 7 and 16 \kms, the   spatial sampling between 
 0.42" and 1.61", and the field-of-view   between 3.6' and 13.7'. The typical total exposure time per  spectral channel is $\sim$4 minutes. 
   
 \vspace*{-0.5cm}
\section{Results of the harmonic decomposition and analysis}
\begin{figure}[h]
   \includegraphics[width=\columnwidth]{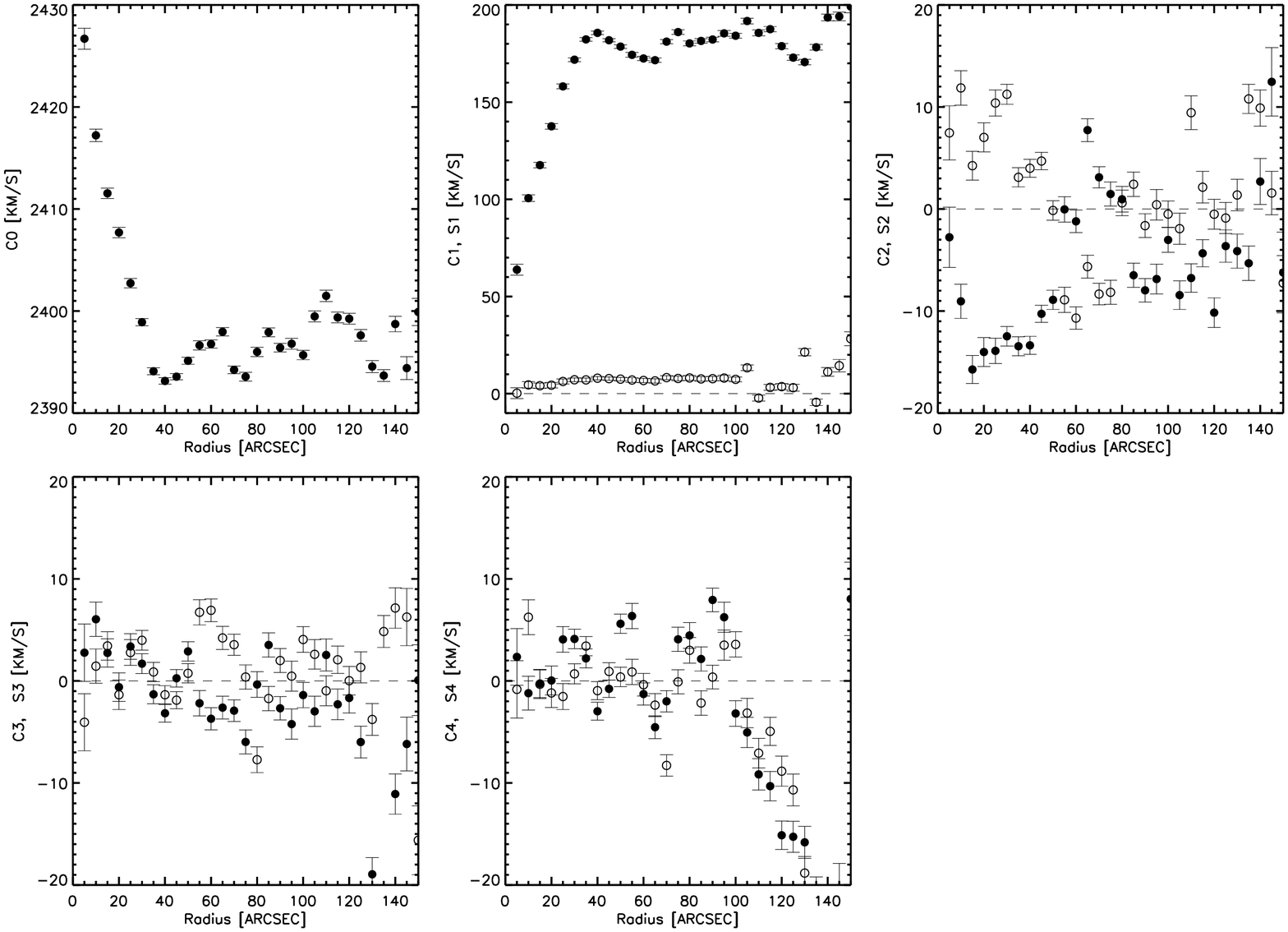}\vspace*{-4.05cm}
  \hspace*{7.7cm}\includegraphics[height=4.0cm]{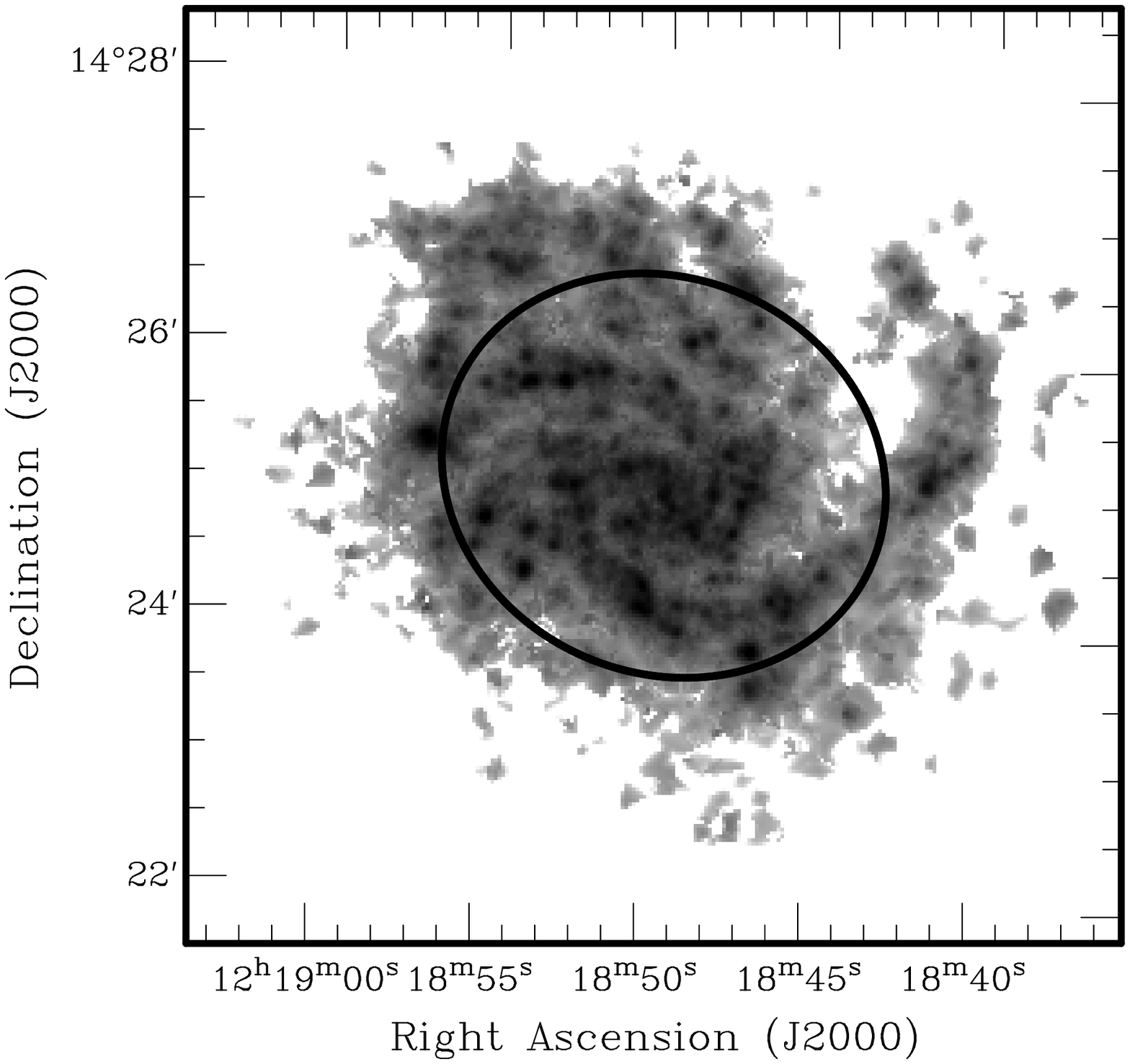} 
  \caption{ Harmonic coefficients $c_m$ and $s_m$ (full and open symbols resp.) up to $m=4$. 
  The ellipse in the \ha\ image displays the projection of the $R=100"$ annulus.}
 \label{fig2}
 \end{figure}

A tilted-ring model is first fitted to the velocity field to derive the inclination, position angle, systemic velocity and
kinematical centre. The velocity field is then expanded into Fourier
coefficients by fitting  $v_{\rm obs} = c_0 + \sum_m{c_m \cos(m\Psi) + s_m \sin(m\Psi)}$ (Schoenmakers et al. 1997).
$\Psi$ corresponds to the angle in the plane 
of a ring, the coefficient  $c_0$  to the systemic velocity of a ring, the first order term  $c_1$ 
to the rotation curve  and all other terms  to non-circular motions.   
Since no warping of the optical disk is clearly detected in this galaxy,  
the inclination ($i$) and kinematic position angle ($P.A.$) can be kept constant as a function of radius during the fitting. The Fourier coefficients are computed up to the 4-th order.  
Figure~\ref{fig2} shows the results of the decomposition inside a radius of $R =150"$. The results are :
\begin{itemize}
\item[$\bullet$] a large variation of the c$_0$ term at small radius which is accompanied by significant non-zero c$_2$ and s$_2$ terms. 
This likely indicates the effect of a lopsided potential ($m = 1$ perturbation). 
\item[$\bullet$] a nearly constant value of 7 $\pm$ 2 \kms\ inside 100" for the s$_1$ term. This feature does not disappear when 
$i$ and/or $P.A.$ are allowed to vary. It could be due to elliptical streamings in a $m=2$ perturbing potential and/or 
to a radial inflow (considering trailing spiral arms). 
\item[$\bullet$]   large variations of the c$_4$ and s$_4$ terms at large radius ($R > 100"$). At these 
radii, the emission is dominated by many \hii\ regions in the northern and western arms and no
evident $m=3$ or $m=5$ modes are detected. The origin of these asymmetries still remains to be explained.     
 \end{itemize} 
 \vspace*{-0.5cm}

This work is in progress and we plan to investigate which event has created the perturbations of the potential 
of NGC 4254. It could be due to the accretion of gas into the disk plane (Phookun, Vogel \& Mundy 1993).


\vspace*{-0.5cm}
\begin{thebibliography}{}
\vspace*{-0.2cm}
\bibitem{} Bournaud F., \& Combes F. 2002, A\&A, 392, 83
\bibitem{} Bournaud  F., et al., 2005, A\&A, 438, 507
\bibitem{} Chemin L., et al., 2005, MNRAS, in press,   astro-ph/0511417 
\bibitem{} Gavazzi G., et al., 2003, A\&A, 400, 451
\bibitem{} Huang S., \& Carlberg R. G., 1997, ApJ, 480, 503
\bibitem{} Moore B., Lake G., Katz N., 1998, ApJ, 499, L5 
\bibitem{} Phookun B., Vogel S. N., \& Mundy L. G., 1993, ApJ, 418, 113
\bibitem{} Rubin V. C., Waterman A. H., \& Kenney J. D. P., 1999, AJ, 118, 236
\bibitem{} Schoenmakers R. H. M., Franx M., \& de Zeeuw P. T., 1997, MNRAS, 292, 349
\bibitem{} Vollmer B., et al., 2001, ApJ, 1561, 708
\end{thebibliography}
\end{document}